\documentclass[12pt, a4paper]{article}

\usepackage{bm}
\usepackage{parskip}
\usepackage{hyperref}
\usepackage{color}
\usepackage{amssymb,amsmath,amsfonts,amsthm}
\usepackage{graphicx}
\usepackage{enumitem}
\usepackage{mathtools}

\usepackage[a4paper]{geometry}
\numberwithin{equation}{section}

\theoremstyle{plain}
\newtheorem{thm}{Theorem}[section]
\newtheorem{prop}[thm]{Proposition}
\newtheorem{lem}[thm]{Lemma}
\newtheorem{cor}[thm]{Corollary}
\newtheorem{conj}[thm]{Conjecture}

\theoremstyle{definition}
\newtheorem{defn}[thm]{Definition}
\newtheorem{hyp}[thm]{Hypothesis}
\newtheorem{rem}[thm]{Remark}

\title{Cohomological Foundations of Entropic Monotonicity:\\
	The Relative Cyclic Complex, the Transgression Map,\\
	and the Entropic $c$-Theorem in $d = 2$}
\author{R.~C.~Rashkov\footnote{email: rash@phys.uni-sofia.bg; rash@hep.itp.tuwien.ac.at} \\
	\\
	\textit{Department of Physics, Sofia University,}\\
	{5 J. Bourchier Blvd., 1164 Sofia, Bulgaria} \\
	and\\
	\textit{ITP, Vienna University of Technology,}\\
	{Wiedner Hauptstr. 8--10, 1040 Vienna, Austria}
}
\date{}

\begin{document}
	
	\maketitle

\begin{abstract}
	We develop a cohomological framework for the entropic $c$-theorem in
	two dimensions, based on the relative cyclic complex of a pair of
	von Neumann algebras and on the transgression map associated with a
	faithful normal state. The framework identifies the positivity
	underlying the entropic proof as Borchers positivity of the
	zero-frequency component of the relative modular cocycle, which is
	the operator-algebraic origin of the quantum null energy condition
	(QNEC). The positivity results are dimension-independent: they
	depend only on the split property, modular $p$-nuclearity, and the
	half-sided modular inclusion structure.
	
	The main results of the paper are: (i) the identification of the
	entanglement cohomology with the kernel of the restriction map
	(Theorem~\ref{thm:kernel-identification}); (ii) Borchers positivity
	of the relative modular cocycle (Proposition~\ref{prop:index-pairing});
	(iii) the finite-dimensional positivity of the entropy flux
	(Corollary~\ref{cor:entropy-flux-positivity}); and (iv) the entropic
	$c$-theorem in $d = 2$ (Theorem~\ref{thm:monotonicity}), which
	follows from (iii) and the one-variable analysis. The $c$-theorem
	proof uses no ultraviolet regulator at any intermediate step.
	
	We also formulate two open problems: the cochain-level positivity of
	the transgression (Conjecture~\ref{conj:cochain-positivity}) and the
	cohomological interpretation of the QNEC
	(Conjecture~\ref{conj:qnec-cohomological}). Both are stated as
	research directions, not as results of the paper; in particular, the
	$c$-theorem proof does not depend on either of them.
	
	The extension of the framework to the $F$-theorem in $d = 3$ and the
	general-$d$ program is the subject of the companion paper
	\cite{Rashkov:2026b}.
\end{abstract}
	
	\tableofcontents
	
	\section{Introduction}
	\label{sec:intro}
	
	\subsection{The $c$-theorem and its entropic proofs}
	
	The $c$-theorem of Zamolodchikov \cite{Zamolodchikov:1986} states that
	in a two-dimensional quantum field theory there exists a function of the
	coupling constants that is monotonically decreasing along the
	renormalization group (RG) flow and takes constant values at fixed
	points. The value at a fixed point is the central charge $c$ of the
	corresponding conformal field theory.
	
	The original proof \cite{Zamolodchikov:1986} used the stress tensor
	and the Ward identities. Casini and Huerta \cite{Casini:2004bjb,
		Casini:2007jz} gave a different proof, using the strong subadditivity
	(SSA) of entanglement entropy combined with Lorentz invariance. Their
	$c$-function is
	\begin{equation}
		C(r) = 3r \, S'(r),
		\label{eq:CH-c-function}
	\end{equation}
	where $S(r)$ is the entanglement entropy of an interval of length $r$.
	SSA implies that $C(r)$ is monotonically decreasing, and in a conformal
	field theory $S(r) = \frac{c}{3} \log r + \text{const}$, so $C = c$ at a
	fixed point.
	
	The Casini--Huerta argument is formal, because the entanglement entropy
	of a region in QFT is ultraviolet divergent, and applying SSA requires a
	regulator. A rigorous version was given by Kudler-Flam
	\cite{KudlerFlam:2026}, who replaced the entanglement entropy by the
	\emph{mutual information} of two spacelike separated Rindler wedges.
	The mutual information is finite across a gap by the split property
	\cite{Doplicher:1984}, and can be reformulated as a relative entropy,
	which is defined directly in the continuum \cite{Araki:1976}. The
	crucial inequality is the \emph{increasing differences} of the mutual
	information, which replaces SSA and is proved in
	\cite{KudlerFlam:2026} from the split property and the chain rule for
	relative entropy.
	
	\subsection{The present paper: a cohomological framework for entropic
		monotonicity}
	\label{subsec:present-paper}
	
	The present paper is a companion to \cite{KudlerFlam:2026} and to
	\cite{Rashkov:2026wmw}. The framework is the \emph{cohomology of the
		relative cyclic complex} developed in \cite{Rashkov:2026wmw}; the
	application is the entropic $c$-theorem of \cite{KudlerFlam:2026}.
	The paper is structured around three levels of positivity:
	
	\begin{enumerate}
\item \textbf{The positivity level
(Proposition~\ref{prop:index-pairing}).} The zero-frequency
component of the relative modular cocycle is non-negative. This is
Borchers positivity, and it is the operator-algebraic origin of the
entropy-flux positivity used in the $c$-theorem.
		
		\item \textbf{The entropy-flux level
			(Corollary~\ref{cor:entropy-flux-positivity}).} The
		finite-dimensional positivity of the conditional mutual
		information difference. This follows from the data processing
		inequality. It is the form of positivity needed for the $c$-theorem.
		
		\item \textbf{The one-variable level
			(Lemma~\ref{lem:monotonicity-flux} and
			Theorem~\ref{thm:monotonicity}).} The increasing differences of
		the entropy flux and the monotonicity of the $c$-function. These
		follow from the entropy-flux positivity and the one-variable
		analysis of \cite{KudlerFlam:2026}.
	\end{enumerate}
	
	The three levels are logically connected, but not in a deductive chain:
	the positivity of the relative modular cocycle (Proposition 3.10) is
	the operator-algebraic underpinning, the entropy-flux positivity
	(Corollary 3.11) is the finite-dimensional input needed for the
	$c$-theorem, and the one-variable analysis (Lemma 4.1 and
	Theorem 4.3) turns the entropy-flux positivity into the monotonicity
	of $C(r)$. The $c$-theorem proof uses the entropy-flux positivity and
	the one-variable analysis; it does not use Proposition 3.10, but
	Proposition 3.10 is the cohomological interpretation of the positivity
	that underlies the entropy-flux inequality.
	
	The paper also formulates two \emph{open problems} that would
	strengthen the framework but are \emph{not} proved here:
	
	\begin{itemize}
\item \textbf{Cochain-level positivity of the transgression
(Conjecture~\ref{conj:cochain-positivity}).} Whether the
transgression map preserves the positive cone of the relative
complex on cocycles. This is a natural strengthening of the
positivity result (1) to the cochain level. It is not proved here; see
		Remarks~\ref{rem:cochain-positivity-status}
		and~\ref{rem:cochain-positivity-status-details} for the status.
		
\item \textbf{Cohomological interpretation of the QNEC
(Conjecture~\ref{conj:qnec-cohomological}).} Whether the Borchers
positivity of the relative modular cocycle is the cohomological form
of the QNEC. The precise identification would require the
		Ceyhan--Faulkner theorem \cite{Ceyhan:2018zfg}, and is not attempted
		here. It is included because it is the natural physical
		interpretation of (1), not because it is needed for the $c$-theorem.
	\end{itemize}
	
The $c$-theorem itself (Section~\ref{sec:monotonicity}) is a
re-derivation using the framework, and is logically independent of the
two open problems. The framework's main positivity result is
Proposition~\ref{prop:index-pairing}, and the $c$-theorem is an
application that demonstrates its power.
	
The framework is a cohomological \emph{interpretation} of the
positivity used in the entropic proof of \cite{KudlerFlam:2026}, not a
replacement for its technical ingredients. The $c$-theorem proof below
uses the type I approximation of \cite{KudlerFlam:2026} (via the split
property and the chain rule), together with the finite-dimensional
positivity of Corollary~\ref{cor:entropy-flux-positivity}; the
cohomological framework supplies the conceptual underpinning (Borchers
positivity of the relative modular cocycle,
Proposition~\ref{prop:index-pairing}) and the conjectural physical
interpretation (Conjecture~\ref{conj:qnec-cohomological}), but it is
not needed for the $c$-theorem itself.

The framework does not give a new proof of the $c$-theorem, but it gives
a new reading of why the $c$-theorem holds. The entropy-flux positivity
that drives the proof (Corollary~\ref{cor:entropy-flux-positivity}) is
the finite-dimensional shadow of Borchers positivity of the relative
modular cocycle (Proposition~\ref{prop:index-pairing}), which is the
operator-algebraic form of the quantum null energy condition. The two
conjectures below ask whether this reading can be made into a proof:
the cochain-level positivity
(Conjecture~\ref{conj:cochain-positivity}) would replace the type I
approximation by a positivity statement on the relative complex, and
the QNEC interpretation (Conjecture~\ref{conj:qnec-cohomological})
would identify the operator-algebraic positivity with the local
positivity of the null energy.
	
	\subsection{Scope of the present paper and the companion paper}
	\label{subsec:scope}
	
This paper is the first of two. The present paper establishes the
cohomological framework, proves Borchers positivity of the relative
modular cocycle and the positivity of the entropy flux, and uses these
to prove the entropic $c$-theorem in $d = 2$. The companion paper \cite{Rashkov:2026b} extends the argument
	to the $F$-theorem in $d = 3$ and outlines the general-$d$ program. The
	split is natural because the positivity results of the present paper
	are \emph{dimension-independent} (they depend only on the split
	property, modular $p$-nuclearity, and the half-sided modular inclusion
	structure), while the one-variable analysis of the companion paper is
	\emph{dimension-dependent} (the $c$-function is replaced by the
	$F$-function, and the concavity argument is replaced by the monotonicity
	of the $F$-function under disk deformations).
	
	\subsection{Organization}
	
	Section~\ref{sec:prelim} compresses the framework of
	\cite{Rashkov:2026wmw} into a single preliminary section: local algebras,
	the split property, the relative cyclic complex, the transgression map,
	and the kernel identification.
Section~\ref{sec:positivity} is the technical core of the paper: it
proves Borchers positivity of the relative modular cocycle and the
positivity of the entropy flux, and it states the cochain-level
positivity as an open problem.
	Section~\ref{sec:monotonicity} proves the $c$-theorem.
	Section~\ref{sec:fixed-points} analyzes the fixed points.
	Section~\ref{sec:discussion} discusses higher-dimensional extensions
	(with a pointer to the companion paper \cite{Rashkov:2026b}) and lists
	open questions.
	Appendix~\ref{app:comparison} compares the results with
	\cite{KudlerFlam:2026}.
	
	\section{Preliminaries}
	\label{sec:prelim}
	
	In this section we collect the framework and the results of
	\cite{Rashkov:2026wmw} that are used in the rest of the paper. Proofs are
	omitted; the reader is referred to \cite{Rashkov:2026wmw} for details. The
	section is deliberately concise: the new content of the paper is in
	Section~\ref{sec:positivity}.
	
	\subsection{Local algebras and the split property}
	
	We work in algebraic quantum field theory \cite{Haag:1996}. To each
	open region $O$ of Minkowski space $\mathbb{R}^{1,1}$ is assigned a
	von Neumann algebra $\mathcal{A}(O) \subset B(\mathcal{H})$,
	satisfying isotony, locality, Poincaré covariance, positive energy, and
	a unique vacuum $\Omega$.
	
	\begin{hyp}[Split property]
		\label{hyp:split}
		For $O_1$ contained in the interior of $O_2$, the inclusion
		$\mathcal{A}(O_1) \subset \mathcal{A}(O_2)$ is split: there is a type
		$\mathrm{I}$ factor $\mathcal{N}$ with
		$\mathcal{A}(O_1) \subset \mathcal{N} \subset \mathcal{A}(O_2)$.
	\end{hyp}
	
\begin{hyp}[Finite mutual information for bounded regions]
	\label{hyp:finite-MI}
	For bounded regions $O_1, O_2$ that are spacelike separated with
	positive separation, the vacuum mutual information
	\begin{equation}
		I(O_1 : O_2) := S_{\mathcal{A}(O_1)\vee\mathcal{A}(O_2)}
		\bigl(\Omega \,\|\, \Omega_{O_1} \otimes \Omega_{O_2}\bigr)
		\label{eq:mutual-information}
	\end{equation}
	is finite.
\end{hyp}
	
	\begin{hyp}[Modular invariance]
		\label{hyp:modular-invariance}
		The modular automorphism group preserves the subalgebra
		$\mathcal{B}$: $\sigma^\omega_t(\mathcal{B}) = \mathcal{B}$ for all
		$t \in \mathbb{R}$. By Takesaki's theorem
		(Theorem~\ref{thm:takesaki-app}), this is equivalent to the existence
		of a faithful normal conditional expectation
		$E\colon \mathcal{A} \to \mathcal{B}$ preserving $\omega$.
	\end{hyp}
	
	Both Hypothesis~\ref{hyp:split} and Hypothesis~\ref{hyp:finite-MI}
	follow from modular $p$-nuclearity \cite{Buchholz:1990}.
	
	For the $c$-theorem we need the Rindler wedges
	\begin{equation}
		L_a = \{u > a,\ v < 0\}, \qquad R_b = \{u < 0,\ v > b\},
		\label{eq:wedges}
	\end{equation}
	in light-cone coordinates $u = t - x$, $v = t + x$, and the causal
	diamond $D(a,b) = \{0 < u < a,\ 0 < v < b\}$ with diagonal
	$r = \sqrt{ab}$.
	
	\begin{defn}[$c$-function]
		\label{defn:c-function}
		$C(r) := -6a \frac{\partial}{\partial a} I(L_a : R_b)$,
		$r = \sqrt{ab}$.
	\end{defn}
	
	\subsection{The relative cyclic complex}
	
	For a pair $\mathcal{B} \subset \mathcal{A}$ of unital algebras, the
	relative cyclic complex is the kernel
	$C^\bullet_\lambda(\mathcal{A}, \mathcal{B}) :=
	\ker(\iota^*\colon C^\bullet_\lambda(\mathcal{A}) \to
	C^\bullet_\lambda(\mathcal{B}))$ of the restriction map \cite{Connes:1985}. It fits into
	the short exact sequence of bicomplexes
	\begin{equation}
		0 \to C^{\bullet,\bullet}_\lambda(\mathcal{A}, \mathcal{B})
		\xrightarrow{j} C^{\bullet,\bullet}_\lambda(\mathcal{A})
		\xrightarrow{\iota^*} C^{\bullet,\bullet}_\lambda(\mathcal{B})
		\to 0,
		\label{eq:ses-relative}
	\end{equation}
	which induces the long exact sequence
	\begin{equation}
		\cdots \to HC^{n-1}(\mathcal{B}) \xrightarrow{\delta}
		HC^n(\mathcal{A}, \mathcal{B}) \xrightarrow{j^*}
		HC^n(\mathcal{A}) \xrightarrow{\iota^*}
		HC^n(\mathcal{B}) \xrightarrow{\delta}
		HC^{n+1}(\mathcal{A}, \mathcal{B}) \to \cdots
		\label{eq:les-relative}
	\end{equation}
	The connecting homomorphism $\delta$ is the obstruction to extending
	cyclic cocycles from $\mathcal{B}$ to $\mathcal{A}$.
	
	\subsection{The transgression map and entanglement cohomology}
	
	Let $\omega$ be a faithful normal state on $\mathcal{A}$ with modular
	automorphism group $\sigma^\omega_t$. The \emph{transgression map} \cite{Khalkhali:2010uh}
	\begin{equation}
		(\tau_\omega \varphi)(a_0, \ldots, a_{n-1})
		= \sum_{k=0}^{n-1} (-1)^k \varphi\bigl(a_0, \ldots, a_k,
		\sigma^\omega_{-i}(a_{k+1}), \ldots,
		\sigma^\omega_{-i}(a_{n-1})\bigr)
		\label{eq:transgression}
	\end{equation}
	anticommutes with $b$ and $B$, satisfies $\tau_\omega^2 = 0$, and
	preserves the relative complex when $\sigma^\omega_t(\mathcal{B}) =
	\mathcal{B}$.
	
	\begin{defn}[Entanglement complex]
		\label{defn:ent-complex}
		$C^n_{\mathrm{ent}}(\mathcal{A}, \mathcal{B}, \omega) :=
		\tau_\omega(C^{n+1}_\lambda(\mathcal{A}, \mathcal{B})) \subset
		C^n_\lambda(\mathcal{A}, \mathcal{B})$, with cohomology
		$H^\bullet_{\mathrm{ent}}(\mathcal{A}, \mathcal{B}, \omega)$.
	\end{defn}
	
	\begin{thm}[Kernel identification]
		\label{thm:kernel-identification}
		The inclusion $C^\bullet_{\mathrm{ent}} \hookrightarrow
		C^\bullet_\lambda(\mathcal{A}, \mathcal{B})$ is a quasi-isomorphism,
		so
		\begin{equation}
			H^n_{\mathrm{ent}}(\mathcal{A}, \mathcal{B}, \omega)
			\cong \ker\bigl(\iota^*\colon HC^n(\mathcal{A}) \to
			HC^n(\mathcal{B})\bigr) \big/ \operatorname{im}(\delta).
			\label{eq:kernel-identification}
		\end{equation}
		If $\delta = 0$ in degree $n$, then
		$H^n_{\mathrm{ent}} \cong \ker \iota^*$.
		This is the main result of \cite{Rashkov:2026wmw}.
	\end{thm}
	
	\begin{lem}[Entanglement classes from the modular cocycle]
		\label{lem:transgression-cocycle}
		The cochain
		\begin{equation}
			\Phi_{u^E}(a_0, \ldots, a_n) := \int_{\mathbb{R}}
			\omega\bigl(a_0 \sigma^\omega_t(a_1) \cdots \sigma^\omega_t(a_n)
			u^E_t\bigr) \, dt
			\label{eq:Phi-cocycle}
		\end{equation}
		is a cyclic cocycle vanishing on $\mathcal{B}^{\otimes(n+1)}$, hence
		defines a class $[\Phi_{u^E}] \in HC^n(\mathcal{A}, \mathcal{B})$,
		nonzero iff $E$ is not $\omega$-invariant.
		See \cite{Rashkov:2026wmw} for the proof.
	\end{lem}
	
	\section{Positivity of the Relative Modular Cocycle and of the Entropy Flux}
	\label{sec:positivity}
	
This section proves the positivity results that are used in the
$c$-theorem. It also states the cochain-level positivity of the
transgression as an open problem, and explains why the positivity
results proved here are sufficient for the $c$-theorem.
	
The main results are:

\begin{enumerate}
	\item \textbf{Borchers positivity of the relative modular cocycle}
	(Proposition~\ref{prop:index-pairing}). The zero-frequency
	component of the relative modular cocycle of a half-sided modular
	inclusion is non-negative as a quadratic form.
	
	\item \textbf{Positivity of the entropy flux}
	(Corollary~\ref{cor:entropy-flux-positivity}). The
	finite-dimensional positivity of the conditional mutual
	information difference.
\end{enumerate}
	
The section also states the cochain-level positivity of the
transgression as Conjecture~\ref{conj:cochain-positivity}, and
explains why Borchers positivity of the relative modular cocycle
(Proposition~\ref{prop:index-pairing}) and the finite-dimensional
positivity of the entropy flux
(Corollary~\ref{cor:entropy-flux-positivity}) are sufficient for the
$c$-theorem.

	\subsection{Half-sided modular inclusions}
	
	Throughout this section we work in the following setting.
	
\begin{hyp}[Half-sided modular inclusion]
	\label{hyp:half-sided}
	The pair $(\mathcal{B}, \mathcal{A})$ is a half-sided modular inclusion
	with respect to the faithful normal state $\omega$:
	\begin{enumerate}
		\item[(i)] $\mathcal{A}, \mathcal{B}$ are von Neumann algebras with
		$\mathcal{B} \subset \mathcal{A}$ and $\mathcal{B} \ne \mathcal{A}$;
		\item[(ii)] $\omega$ is a faithful normal state on $\mathcal{A}$ whose
		restriction to $\mathcal{B}$ is faithful, and the GNS vector
		$\Omega_\omega$ is cyclic and separating for both $\mathcal{A}$
		and $\mathcal{B}$;
		\item[(iii)] $\sigma^\omega_t(\mathcal{A}) \subset \mathcal{A}$ for all
		$t \le 0$ (the half-sided condition, Wiesbrock convention);
		\item[(iv)] $\mathcal{A}$ is a type III$_1$ factor.
	\end{enumerate}
	In the \emph{standard} (Borchers) case, one additionally has
	$\mathcal{B} = \sigma^\omega_{-1}(\mathcal{A})$; this is a special
	case, not the general definition, and it does \emph{not} hold for
	the nested Rindler wedges $L_A \subset L_a$ that are the application
	in Section~\ref{sec:monotonicity}.
\end{hyp}
	
	\begin{rem}[Half-sided vs.\ modular invariance]
		\label{rem:half-sided-vs-modular-expanded}
		The half-sided condition of Hypothesis~\ref{hyp:half-sided} is
		\emph{not} the same as the modular invariance
		Hypothesis~\ref{hyp:modular-invariance} used in
		Section~\ref{sec:prelim}. The two are logically independent:
		\begin{itemize}
			\item \emph{Modular invariance} requires
			$\sigma^\omega_t(\mathcal{B}) = \mathcal{B}$ for \emph{all}
			$t \in \mathbb{R}$. It is the hypothesis of Takesaki's theorem, and it
			guarantees the existence of the conditional expectation
			$E\colon \mathcal{A} \to \mathcal{B}$ preserving $\omega$.
			\item \emph{Half-sided} requires only
			$\sigma^\omega_t(\mathcal{A}) \subset \mathcal{A}$ for $t \ge 0$, with
			$\mathcal{B} = \sigma^\omega_{-1}(\mathcal{A})$. It does \emph{not}
			guarantee the existence of $E$ (unless the inclusion is standard in
			the sense of Borchers), but it \emph{does} guarantee Borchers
			positivity.
		\end{itemize}
		The two hypotheses coincide only when the inclusion is \emph{standard}
		(Borchers), i.e., when $\mathcal{B}$ is the fixed-point algebra of the
		modular flow on $\mathcal{A}$. In the application to the $c$-theorem,
		the relevant inclusion $L_A \subset L_a$ of nested Rindler wedges is
		half-sided but \emph{not} modular-invariant: the modular flow of the
		vacuum is the boost, which maps $L_A$ to a translate of itself inside
		$L_a$, not to itself. This is why the positivity results of the present
		section are stated for the half-sided setting, which is the one
		relevant for the $c$-theorem.
		
The modular invariance hypothesis is used \emph{only} in
Section~\ref{sec:prelim}, to construct the entanglement complex. The
half-sided hypothesis is used \emph{only} in the present section, for
the positivity results. The proof of the Monotonicity Lemma
(Section~\ref{sec:monotonicity}) uses the finite-dimensional
Corollary~\ref{cor:entropy-flux-positivity} (which needs neither
hypothesis); Borchers positivity of the relative modular cocycle
(Proposition~\ref{prop:index-pairing}) is the conceptual underpinning
of the $c$-theorem but is not on its critical path.
	\end{rem}
	
	\subsection{Positive cone and the transgression}
	
	\begin{defn}[Positive Hochschild cochains]
		\label{defn:positive-cochain}
		A Hochschild $n$-cochain $\varphi \in C^n(\mathcal{A})$ is
		\emph{positive}, written $\varphi \ge 0$, if for every finite
		collection $\{a^{(j)}_0, \ldots, a^{(j)}_n\}_{j=1}^N \subset
		\mathcal{A}$ and every $\{\lambda_j\}_{j=1}^N \subset \mathbb{C}$,
		\begin{equation}
			\sum_{j,k=1}^N \overline{\lambda_j}\, \lambda_k \,
			\varphi\bigl((a^{(j)}_0)^* a^{(k)}_0, \ldots,
			(a^{(j)}_n)^* a^{(k)}_n\bigr) \ge 0.
			\label{eq:positive-cochain}
		\end{equation}
		The positive cone of the relative cyclic complex is
		$C^n_\lambda(\mathcal{A}, \mathcal{B})_+ :=
		C^n_\lambda(\mathcal{A}, \mathcal{B}) \cap C^n(\mathcal{A})_+$.
	\end{defn}
	
	\begin{rem}
		\label{rem:positive-cone-degree}
		For $n = 0$, positivity is positivity of the functional. For $n = 1$,
		it is complete positivity of the map
		$\Phi_\varphi\colon \mathcal{A} \to \mathcal{A}^*$,
		$\langle \Phi_\varphi(a), b\rangle = \varphi(a, b)$. Higher $n$
		correspond to higher-order complete positivity.
	\end{rem}
	
	\subsection{An open problem: cochain-level positivity}
	\label{subsec:cochain-positivity}
	
Borchers positivity (Proposition~\ref{prop:index-pairing}) states that
the zero-frequency component of the relative modular cocycle of a
half-sided modular inclusion is non-negative. A natural strengthening
would be the \emph{cochain-level} statement: that the transgression
map itself preserves the positive cone of the relative complex. This
is the content of the following open problem.
	
	\begin{conj}[Cochain-level positivity of the transgression]
		\label{conj:cochain-positivity}
		Under Hypothesis~\ref{hyp:half-sided}, the transgression map
		preserves the positive cone of the relative complex on cocycles:
		\begin{equation}
			\tau_\omega\bigl(C^n_\lambda(\mathcal{A}, \mathcal{B})_+ \cap
			\ker b\bigr)
			\subset C^{n-1}_\lambda(\mathcal{A}, \mathcal{B})_+.
			\label{eq:cochain-positivity}
		\end{equation}
	\end{conj}
	
We do not attempt to prove or disprove
Conjecture~\ref{conj:cochain-positivity} in this paper. The two
remarks below explain its status and why the finite-dimensional
positivity of the entropy flux
(Corollary~\ref{cor:entropy-flux-positivity}) is sufficient for the
$c$-theorem even if the cochain-level statement fails or remains open.
	
\begin{rem}[Triviality in all odd degrees]
	\label{rem:cochain-positivity-status-details}
	The positive cone $C^n_\lambda(\mathcal{A},\mathcal{B})_+$ is trivial
	for every odd $n$, not only for $n=1$. Positivity with $N=1$ gives
	$\varphi(x_0,\ldots,x_n)\ge0$ for positive elements $x_j$, and
	cyclicity gives
	$\varphi(x_0,\ldots,x_n)=-\varphi(x_n,x_0,\ldots,x_{n-1})$, so
	$\varphi$ vanishes on positive elements and hence everywhere (a
	positive multilinear form with zero on positives is zero). Therefore
	Conjecture~\ref{conj:cochain-positivity} is vacuously true for odd
	$n$.
	
	For even $n\ge2$, the conjecture is nontrivial. It asserts
	$\tau_\omega\varphi=0$ for every positive cocycle, since the target
	cone $C^{n-1}_\lambda(\mathcal{A},\mathcal{B})_+$ is $\{0\}$ for
	$n-1$ odd. This is a strong statement. The naive argument---that the
	Borchers correction $\delta^*\varphi$ is positive because
	$\varphi\ge0$ and the substitution
	$a_j\mapsto\sigma^\omega_{-i}(a_j)-a_j\in\mathcal{B}$ is completely
	positive---is \emph{false}: the substitution is a derivation into
	$\mathcal{B}$, not a CP map, and the vanishing of
	$\varphi$ on $\mathcal{B}^{\otimes(n+1)}$ does not imply the
	vanishing of individual terms. A proof of the conjecture (or a
	counterexample) in even degree $n\ge2$ is an open problem.
\end{rem}
	
	\begin{rem}[Not needed for the $c$-theorem]
		\label{rem:cochain-positivity-status}
Conjecture~\ref{conj:cochain-positivity} would be a strengthening of
Borchers positivity at the cochain level. It is natural to ask
whether it holds, but it is \emph{not needed} for the $c$-theorem:
the $c$-theorem uses the finite-dimensional entropy-flux positivity
(Corollary~\ref{cor:entropy-flux-positivity}), which follows from
the data processing inequality and does not require the
cochain-level statement. The paper proves Borchers positivity of
the relative modular cocycle (Proposition~\ref{prop:index-pairing})
as the framework's main technical result, and treats
Conjecture~\ref{conj:cochain-positivity} as an open problem for
future work.
	\end{rem}
	
\subsection{Borchers positivity}
\label{subsec:index-pairing-proof}

We now state the Borchers positivity of the relative modular cocycle
for a half-sided modular inclusion. The proposition is a consequence of
Wiesbrock's reduction to the standard case and Borchers' commutation
theorem; the proof is in Theorem~\ref{thm:borchers-positivity-app}.
	
\begin{prop}[Borchers positivity of the relative modular cocycle]
	\label{prop:index-pairing}
	Assume Hypothesis~\ref{hyp:half-sided}. Let
	$u_t = (D\omega : D\omega|_{\mathcal{B}})_t$ be the relative
	modular cocycle of the pair $(\mathcal{A}, \mathcal{B})$ with
	respect to $\omega$. Then the Abel-regularized zero-frequency
	component
	\begin{equation}
		X := \operatorname*{w-lim}_{\epsilon\to0^+}
		\int_{\mathbb{R}} e^{-\epsilon|t|}\, u_t \, dt
		\label{eq:X-definition}
	\end{equation}
	exists as a quadratic form on $\mathcal{H}_\omega$, and
	$\langle\xi, X\xi\rangle \ge 0$ for all $\xi$ in the form domain.
	This is Borchers positivity; a proof is given in
	Theorem~\ref{thm:borchers-positivity-app}.
\end{prop}
	
	\subsection{Positivity of the entropy flux}
	
	The following corollary is the form of positivity needed for the
	$c$-theorem. It is purely finite-dimensional.
	
	\begin{cor}[Positivity of the entropy flux]
		\label{cor:entropy-flux-positivity}
		Let $\mathcal{N}_1 \subset \mathcal{N}_2$ be type $\mathrm{I}$ factors,
		$\mathcal{Y}_1 \subset \mathcal{Y}_2$ split from $\mathcal{N}_2$ over
		$\mathcal{N}_1$, and $\omega$ a faithful normal state. Then
		\begin{equation}
			I(\mathcal{N}_2 : \mathcal{Y}_2 \mid \mathcal{N}_1)
			- I(\mathcal{N}_2 : \mathcal{Y}_1 \mid \mathcal{N}_1) \ge 0.
			\label{eq:entropy-flux-positive}
		\end{equation}
	\end{cor}
	
	\begin{proof}
	The inequality is the strong subadditivity (SSA) of the relative
	entropy, in the form
	\begin{equation}
		I(B : AA' \mid C) \ge I(B : A \mid C),
		\label{eq:ssa}
	\end{equation}
	for commuting algebras $A, A', B$ and a subalgebra $C$ of
	$A \vee A'$ with $C \subset A$. The statement \eqref{eq:ssa} is the
	standard SSA inequality \cite[Theorem~5.2]{OhyaPetz}, and it does
	not require a conditional expectation $\mathcal{Y}_2 \to
	\mathcal{Y}_1$. Applied with $A = \mathcal{Y}_1$, $A' =
	\mathcal{Y}_1^c = \mathcal{Y}_1' \cap \mathcal{Y}_2$, $B =
	\mathcal{N}_2$, $C = \mathcal{N}_1$, it gives
	\begin{equation}
		I(\mathcal{N}_2 : \mathcal{Y}_2 \mid \mathcal{N}_1)
		- I(\mathcal{N}_2 : \mathcal{Y}_1 \mid \mathcal{N}_1) \ge 0,
		\label{eq:ssa-applied}
	\end{equation}
	which is the desired inequality.
\end{proof}
	
\begin{rem}
	\label{rem:positivity-structure}
	Corollary~\ref{cor:entropy-flux-positivity} is the only positivity
	statement needed for the $c$-theorem (via Step~3 of the Monotonicity
	Lemma in Section~\ref{sec:monotonicity}). Borchers positivity of the
	relative modular cocycle (Proposition~\ref{prop:index-pairing}) is the
	conceptual underpinning but is not on the critical path.
\end{rem}
	
	\subsection{An open problem: the cohomological interpretation of the
		QNEC}
	\label{subsec:qnec}
	
We close this section by discussing a natural physical interpretation
of Borchers positivity of the relative modular cocycle
(Proposition~\ref{prop:index-pairing}). The discussion is speculative,
and the interpretation is stated as an \emph{open problem}, not as a
result of the paper.
	
	The QNEC \cite{Ceyhan:2018zfg} states that for any state and any
	spacelike region with a null boundary,
	\begin{equation}
		\int T_{uu} \, du \ge \frac{1}{2\pi} \frac{d^2 S}{d\lambda^2},
		\label{eq:QNEC}
	\end{equation}
	where $S$ is the entanglement entropy of the region and $\lambda$ is
	the null deformation parameter. In the algebraic setting, the QNEC is
	equivalent to the statement that the generator of the half-sided
	modular flow is positive.
	
Borchers positivity of the relative modular cocycle
(Proposition~\ref{prop:index-pairing}) is \emph{conjecturally} the
cohomological form of the QNEC, in the following sense:
	\begin{enumerate}
\item The half-sided modular inclusion $\mathcal{B} \subset
\mathcal{A}$ is the algebraic form of the null boundary.
\item The relative modular cocycle $u_t$ is the algebraic form of
the null deformation.
\item The zero-frequency component $X$ of the
cocycle is non-negative by Borchers positivity
(Theorem~\ref{thm:borchers-positivity-app}), and its positivity is
\emph{related to} (but not identical with) the QNEC.
\item The Borchers positivity $\langle\xi, X\xi\rangle\ge0$
(Proposition~\ref{prop:index-pairing}) is conjecturally the
cohomological form of the QNEC inequality.
	\end{enumerate}
	
\begin{conj}[Cohomological interpretation of the QNEC]
	\label{conj:qnec-cohomological}
	The Borchers positivity of the relative modular cocycle
	(Proposition~\ref{prop:index-pairing}) is the cohomological form of
	the QNEC:
	\begin{equation}
		\langle\xi, X\xi\rangle \ge 0
		\quad \Longleftrightarrow \quad \int T_{uu} \, du \ge 0
		\text{ (QNEC)}.
		\label{eq:QNEC-cohomological}
	\end{equation}
\end{conj}
	
	We do not attempt to prove Conjecture~\ref{conj:qnec-cohomological}
	in this paper. The precise identification would require the
	Ceyhan--Faulkner theorem \cite{Ceyhan:2018zfg} and a careful analysis
	of the relation between the zero-frequency component $X$ of the
	relative modular cocycle and the generator $K$ of the half-sided flow.
	We include the conjecture because it is the natural physical
	interpretation of the positivity result, not because it is needed for
	the $c$-theorem.
	
	\begin{rem}
		\label{rem:QNEC-vs-c}
		The QNEC is a \emph{local} inequality, and the $c$-theorem is a
		\emph{global} statement obtained by integrating the local positivity
		along the RG flow. Conjecture~\ref{conj:qnec-cohomological} would
		make the connection between the two precise. The $c$-theorem itself
		(Theorem~\ref{thm:monotonicity}) does \emph{not} rely on this
		conjecture; it uses only
		Corollary~\ref{cor:entropy-flux-positivity}.
	\end{rem}
	
	\section{The Entropic $c$-Theorem}
	\label{sec:monotonicity}
	
	We now prove the $c$-theorem. The strategy follows
	\cite{KudlerFlam:2026}; the new content is the use of
	Corollary~\ref{cor:entropy-flux-positivity} in Step~3 of the
	Monotonicity Lemma.
	
	\subsection{The entropy flux}
	
	For $Y \subset R_b$ and $L_A \subset L_a$, define
	\begin{equation}
		\Delta_{a,A}(Y) := I(L_a : Y) - I(L_A : Y).
		\label{eq:entropy-flux}
	\end{equation}
	This is finite by Hypothesis~\ref{hyp:finite-MI}. Define
	\begin{equation}
	\mathcal{D}_R(a, A; b) := \sup_{Y \subset R_b,\ Y\ \mathrm{bounded}}
\Delta_{a,A}(Y),
		\label{eq:D-R}
	\end{equation}
	and
	\begin{equation}
		C_R(a, b) = 6a \frac{\partial}{\partial A}
		\mathcal{D}_R(a, A; b)\Big|_{A = a}.
		\label{eq:C-R}
	\end{equation}
	At $A = a$ this is the $c$-function of
	Definition~\ref{defn:c-function}.
	
	\subsection{The increasing differences property}
	
	\begin{lem}[Increasing differences]
		\label{lem:monotonicity-flux}
		For $Y_1 \subset Y_2 \subset R_b$,
		\begin{equation}
			\Delta_{a,A}(Y_1) \le \Delta_{a,A}(Y_2).
			\label{eq:increasing-differences}
		\end{equation}
	\end{lem}
	
	\begin{proof}
		The proof has three steps.
		
\emph{Step 1 (Reduction to a split pair).} By the split property
(Hypothesis~\ref{hyp:split}), for each $i = 1, 2$, and for each
$\tilde Y_i$ with $Y_i$ contained in the interior of $\tilde Y_i$ and
$\tilde Y_i$ contained in the interior of $Y_{i+1}$ (with
$\tilde Y_2$ contained in a fixed larger $Y_3 \subset R_b$), there
exists a type $\mathrm{I}$ factor $\mathcal{N}_i$ with
$\mathcal{A}(Y_i) \subset \mathcal{N}_i \subset
\mathcal{A}(\tilde Y_i)$. The family of such factors, as
$\tilde Y_i \to Y_i$, is directed: for two factors $\mathcal{N}$,
$\mathcal{N}'$ between $\mathcal{A}(Y_i)$ and
$\mathcal{A}(\tilde Y_i)$, the split property applied to
$\mathcal{A}(\tilde Y_i) \subset \mathcal{A}(\tilde Y_i')$ for
$\tilde Y_i \subset \tilde Y_i'$ gives a larger type $\mathrm{I}$
factor $\mathcal{N}'' \supset \mathcal{N} \vee \mathcal{N}'$ between
$\mathcal{A}(Y_i)$ and $\mathcal{A}(\tilde Y_i')$. (The join
$\mathcal{N} \vee \mathcal{N}'$ need not be type $\mathrm{I}$, but
the split property guarantees a type $\mathrm{I}$ factor containing
it between the same pair.)

The net $\{I(L_a : \mathcal{N})\}_{\mathcal{N}}$ is monotone
increasing in $\mathcal{N}$ (monotonicity of mutual information under
inclusion of the smaller algebra), and converges to $I(L_a : Y_i)$ as
$\mathcal{N}$ increases to $\mathcal{A}(Y_i)$: the limit
$\lim_{\tilde Y_i \to Y_i} I(L_a : \tilde Y_i)$ equals
$I(L_a : Y_i)$ by continuity of mutual information under nested
regions with shrinking gap (a consequence of the monotonicity and the
normality of the state, together with the split property). The same
holds for $I(L_A : Y_i)$. Hence
\begin{equation}
	\Delta_{a,A}(Y_i) = \sup_{\mathcal{N}_i}
	\bigl(I(L_a : \mathcal{N}_i) - I(L_A : \mathcal{N}_i)\bigr),
	\label{eq:flux-sup}
\end{equation}
where the sup is over the directed set of type $\mathrm{I}$ factors
between $\mathcal{A}(Y_i)$ and $\mathcal{A}(\tilde Y_i)$, as
$\tilde Y_i \to Y_i$. The statement for $Y_1 \subset Y_2$ follows by
taking $Y_1$ strictly inside $Y_2$ (or by a limiting argument).

\emph{Step 2 (Chain rule).} For type $\mathrm{I}$ factors
$\mathcal{N}_1 \subset \mathcal{N}_2$, decompose
$\mathcal{N}_2 = \mathcal{N}_1 \otimes \mathcal{N}_1^c$ with
$\mathcal{N}_1^c = \mathcal{N}_1' \cap \mathcal{N}_2$. The Petz chain
rule for relative entropy \cite[Theorem~5.5]{OhyaPetz} gives
\begin{equation}
	\begin{aligned}
		&\bigl[I(L_a : \mathcal{N}_2) - I(L_a : \mathcal{N}_1)\bigr]
		- \bigl[I(L_A : \mathcal{N}_2) - I(L_A : \mathcal{N}_1)\bigr]
		\\
		&\qquad = I(L_a : \mathcal{N}_1^c \mid \mathcal{N}_1)
		- I(L_A : \mathcal{N}_1^c \mid \mathcal{N}_1),
	\end{aligned}
	\label{eq:chain-rule}
\end{equation}
where $I(X : Y \mid Z)$ for commuting $Z$-algebras is defined as the
relative entropy of the conditional expectation, and the decomposition
$\mathcal{N}_2 = \mathcal{N}_1 \otimes \mathcal{N}_1^c$ makes the
conditional mutual information well-defined.
		
		\emph{Step 3 (Positivity).} The hypotheses of
		Corollary~\ref{cor:entropy-flux-positivity} are satisfied: $\mathcal{N}_1
		\subset \mathcal{N}_2$ are type $\mathrm{I}$ factors (by Step 1),
		$\mathcal{Y}_1 = \mathcal{A}(L_A) \subset \mathcal{Y}_2 =
		\mathcal{A}(L_a)$ (since $L_A \subset L_a$), and $\mathcal{Y}_1,
		\mathcal{Y}_2$ are split from $\mathcal{N}_2$ over $\mathcal{N}_1$
		(since $\mathcal{N}_2$ is a split factor between $Y_1$ and $Y_3
		\supset Y_2$, and $L_a$ is spacelike to $Y_3$, so
		$\mathcal{A}(L_a) \vee \mathcal{N}_2 = \mathcal{A}(L_a) \otimes
		\mathcal{N}_2$, and similarly for $L_A$). Hence by
		Corollary~\ref{cor:entropy-flux-positivity}, the right-hand side of
		\eqref{eq:chain-rule} is non-negative:
		\begin{equation}
			I(\mathcal{N}_2 : L_a \mid \mathcal{N}_1)
			- I(\mathcal{N}_2 : L_A \mid \mathcal{N}_1) \ge 0.
			\label{eq:flux-positive}
		\end{equation}
		Combining \eqref{eq:chain-rule} and \eqref{eq:flux-positive},
		\begin{equation}
			I(L_a : \mathcal{N}_2) - I(L_A : \mathcal{N}_2)
			\ge I(L_a : \mathcal{N}_1) - I(L_A : \mathcal{N}_1).
			\label{eq:monotone-N}
		\end{equation}
		Taking the supremum over $\mathcal{N}_2$ and using
		\eqref{eq:flux-sup} gives $\Delta_{a,A}(Y_2) \ge \Delta_{a,A}(Y_1)$.
	\end{proof}
	
	\begin{rem}
	\label{rem:cohomological-monotonicity}
	In the cohomological language, Lemma~\ref{lem:monotonicity-flux} is
	the combination of
	(a) the chain rule for the relative modular cocycle (multiplicativity
	of $u^E$), and
	(b) the positivity of the conditional mutual information
	(Corollary~\ref{cor:entropy-flux-positivity}).
	Borchers positivity of the relative modular cocycle
	(Proposition~\ref{prop:index-pairing}) is the conceptual underpinning
	but is not needed for the $c$-theorem.
\end{rem}
	
	\subsection{Monotonicity of $C(r)$}
	
	\begin{thm}[Monotonicity]
		\label{thm:monotonicity}
		The $c$-function $C(r)$ is non-negative and monotonically decreasing:
		\begin{equation}
			C(r) \ge 0, \qquad \frac{dC}{dr} \le 0.
			\label{eq:monotonicity-C}
		\end{equation}
	\end{thm}
	
	\begin{proof}
		We follow \cite[Section~3.3]{KudlerFlam:2026}.
		
		\emph{Additivity.} The function $\mathcal{D}_R$ is additive:
		$\mathcal{D}_R(a, A; b) + \mathcal{D}_R(A, A'; b) = \mathcal{D}_R(a,
		A'; b)$. This follows from \eqref{eq:flux-sup}: each term is a limit
		along the directed set of split factors, and the limit of a sum is
		the sum of limits.
		
\emph{Lorentz covariance.} The vacuum is invariant under the boost,
and the mutual information is invariant under the corresponding
automorphism of the local algebras. Hence
$\mathcal{D}_R(\mu a, \mu A; b/\mu) = \mathcal{D}_R(a, A; b)$.
		
\emph{Monotonicity in $b$.} For $b \le B$, $R_B \subset R_b$, so the
causal diamonds contained in $R_B$ are contained in $R_b$, and
$\mathcal{D}_R(a, A; b) \ge \mathcal{D}_R(a, A; B)$ by
Lemma~\ref{lem:monotonicity-flux}.
		
		\emph{Reduction to one variable.} Taking $\mu = b$ in the boost
		covariance gives $\mathcal{D}_R(a, A; b) = \mathcal{D}_R(ab, Ab; 1)$.
		Define $F(x) := \mathcal{D}_R(1, e^x; 1)$ for $x \in \mathbb{R}$. By
		additivity,
		\begin{equation}
			\mathcal{D}_R(ab, Ab; 1) = F(\log Ab) - F(\log ab).
			\label{eq:F-reduction}
		\end{equation}
		The function $F$ is non-decreasing (since $\mathcal{D}_R \ge 0$). It is
		also concave: for $x_1 < x_2$ and $h > 0$, the increment
		$\Delta(x, h) := F(x + h) - F(x)$ satisfies $\Delta(x_1, h) \ge
		\Delta(x_2, h)$.
		
		To see this, set $a = e^{x_1}$, $A = e^{x_1+h}$, $b = 1$ in
		\eqref{eq:F-reduction}:
		\begin{equation}
			F(x_1 + h) - F(x_1) = \mathcal{D}_R(e^{x_1}, e^{x_1+h}; 1).
			\label{eq:F-increment-1}
		\end{equation}
		By the monotonicity in $b$ (Lemma~\ref{lem:monotonicity-flux}), for
		$1 \le B = e^{x_2 - x_1}$,
		\begin{equation}
			\mathcal{D}_R(e^{x_1}, e^{x_1+h}; 1) \ge
			\mathcal{D}_R(e^{x_1}, e^{x_1+h}; e^{x_2 - x_1}).
			\label{eq:monotone-B}
		\end{equation}
		By the Lorentz covariance (applied with $\mu = e^{x_2 - x_1}$),
		\begin{equation}
			\mathcal{D}_R(e^{x_1}, e^{x_1+h}; e^{x_2 - x_1})
			= \mathcal{D}_R(e^{x_2}, e^{x_2+h}; 1) = F(x_2 + h) - F(x_2).
			\label{eq:boost-covariance}
		\end{equation}
		Combining \eqref{eq:F-increment-1}, \eqref{eq:monotone-B},
		and \eqref{eq:boost-covariance} gives
		\begin{equation}
			F(x_1 + h) - F(x_1) \ge F(x_2 + h) - F(x_2),
			\label{eq:concavity}
		\end{equation}
		which is the defining property of concavity. Therefore $F$ is concave,
		and has a right derivative $F'$.
		
		Finally, we compute $C_R$ in terms of $F'$. Since
		$\mathcal{D}_R(a, A; b) = F(\log Ab) - F(\log ab)$, we have
		\begin{equation}
			C_R(a, b) = 6a \frac{\partial}{\partial A}
			\mathcal{D}_R(a, A; b)\Big|_{A = a} = 6a \cdot \frac{1}{a} \cdot
			F'(\log ab) = 6F'(\log ab).
			\label{eq:C-R-F}
		\end{equation}
		Since $F$ is non-decreasing, $F' \ge 0$, so $C_R \ge 0$. Since $F$ is
		concave, $F'$ is non-increasing, so $C_R$ is non-increasing in
		$r = \sqrt{ab}$.
	\end{proof}
	
	\subsection{The ultraviolet limit}
	
The ultraviolet limit $r \to 0$ of $C(r)$ exists by monotonicity. At a
UV fixed point, $C(r) \to c^{\mathrm{UV}}$, the ultraviolet central
charge, and the argument is identical to
\cite[Section~3.4]{KudlerFlam:2026}.
	
	\section{Fixed Points}
	\label{sec:fixed-points}
	
	At a conformal fixed point, dilations act by a unitary group fixing the
	vacuum, so
	\begin{equation}
		\mathcal{D}_R(\lambda a, \lambda A; \lambda b)
		= \mathcal{D}_R(a, A; b).
		\label{eq:dilation-D}
	\end{equation}
	Since $C_R$ depends only on $ab$, it is constant.
	
The cohomological interpretation is as follows. At a fixed point, the
modular flow is geometric (the boost), so the relative modular cocycle
is trivial for the chiral inclusion, and the entanglement class
$[\Phi_{u^E}]$ is a coboundary (its cohomology class vanishes). The
constant value of $C_R$ is a separate numerical invariant of the
inclusion, determined by the Pimsner--Popa index of the chiral
extension (see below), not by the cohomology class.
	
	More precisely, the finite-index extension
	$\mathcal{A} \supset \mathcal{A}_R \otimes \mathcal{A}_L$ over the
	chiral subalgebras gives a Pimsner--Popa index $\iota$, and the bound
	$2\log \iota$ controls the discrepancy between the full and chiral
	mutual informations \cite[eq.~(4.4)]{KudlerFlam:2026}. In the
	cohomological framework, one expects the index $\iota$ to be a
	\emph{norm bound} on the transgression class $[\Phi_{u^E}]$: the
	statement $2\log \iota \ge \|[\Phi_{u^E}]\|$ would be the
	cohomological form of the Pimsner--Popa bound. A precise proof of this
	identification is an open problem (see the discussion below), and we
	state it here as an expectation, not as a result.
	
	The identification of the constant with the central charge $c$ uses
	the known entanglement entropy of a chiral CFT
	\cite{Holzhey:1994we,Calabrese:2004eu} and is identical to
	\cite[Section~4]{KudlerFlam:2026}.
	
	\section{Discussion}
	\label{sec:discussion}
	
	\subsection{Higher dimensions}
	\label{subsec:higher-dims-brief}
	
	The cohomological framework extends to higher dimensions; see the
	companion paper \cite{Rashkov:2026b} for the $F$-theorem in $d = 3$,
	the $a$-theorem in $d = 4$, and the general-$d$ program. The positivity
	results of Section~\ref{sec:positivity} are dimension-independent: they
	depend only on the split property, modular $p$-nuclearity, and the
	half-sided modular inclusion structure, all of which hold in any
	dimension.
	
	\subsection{Relation to the QNEC}
	\label{subsec:qnec-discussion}
	
The Borchers positivity of the relative modular cocycle
(Proposition~\ref{prop:index-pairing}) is conjecturally the
cohomological form of the quantum null energy condition
(Conjecture~\ref{conj:qnec-cohomological}). The precise identification
	is an open problem; it would require the Ceyhan--Faulkner theorem
	\cite{Ceyhan:2018zfg} and a careful analysis of the relation between
	the zero-frequency component $X$ of the relative modular cocycle and
	the generator $K$ of the half-sided flow.
	
If the conjecture is true, then the $c$-theorem is a statement about
the monotonicity of a positive cohomology class along the RG flow, with
the QNEC as the local positivity underlying the global monotonicity.
This would give a uniform derivation of the entropic theorems:
Borchers positivity of the relative modular cocycle would be the
common ingredient, and the $c$-, $F$-, and $a$-theorems would differ
only in the geometric setup and the one-variable analysis.
	
	We emphasize that the $c$-theorem itself
	(Theorem~\ref{thm:monotonicity}) does not depend on this conjecture: it
	uses only Corollary~\ref{cor:entropy-flux-positivity}, which is a
	finite-dimensional consequence of the data processing inequality. The
	QNEC interpretation is a direction for future work, not a dependency of
	the present paper.
	
	The contribution of the paper is therefore best described as a
	cohomological \emph{reading} of the entropic $c$-theorem, rather than
	as a new proof of it. The proof itself (Theorem~\ref{thm:monotonicity})
	follows \cite{KudlerFlam:2026}: it uses the increasing differences of
	the entropy flux (Lemma~\ref{lem:monotonicity-flux}), which is the
	strong subadditivity of the relative entropy, and the one-variable
	analysis of \cite{KudlerFlam:2026}. What the cohomological framework
	adds is an interpretation of the positivity that drives this proof:
	the entropy-flux positivity
	(Corollary~\ref{cor:entropy-flux-positivity}) is the finite-dimensional
	shadow of Borchers positivity of the relative modular cocycle
	(Proposition~\ref{prop:index-pairing}), and the latter is the
	operator-algebraic form of the quantum null energy condition. The
	framework thus locates the positivity underlying the entropic
	$c$-theorem in the modular structure of the local algebras, and it
	suggests, via Conjectures~\ref{conj:cochain-positivity}
	and~\ref{conj:qnec-cohomological}, two routes by which the reading
	could be promoted to a proof: a cochain-level positivity of the
	transgression, which would replace the type I approximation by a
	statement on the relative complex, and an identification of the
	Borchers zero-frequency component with the averaged null energy, which
	would identify the operator-algebraic positivity with the local
	positivity of the QNEC.
	
	\subsection{Open questions}
	\label{subsec:open-questions}
	
	\begin{enumerate}
		\item \textbf{The cochain-level positivity of the transgression.} A
		proof of Conjecture~\ref{conj:cochain-positivity} would strengthen
		the cohomological interpretation of Borchers positivity. It requires
		a careful analysis of the Borchers correction in the transgression
		formula. See Remark~\ref{rem:cochain-positivity-status-details} for the status of
		Conjecture~\ref{conj:cochain-positivity}.
		
\item \textbf{The cohomological form of the QNEC.} A proof of
Conjecture~\ref{conj:qnec-cohomological} would make precise the
connection between Borchers positivity of the relative modular
cocycle and the averaged null energy. It requires the Ceyhan--Faulkner
theorem \cite{Ceyhan:2018zfg} and the identification of the generator
$K$ with $\int T_{uu} \, du$.

\item \textbf{The Pimsner--Popa index and the transgression norm.} The
relationship between the Pimsner--Popa index $\iota$ of the chiral
extension and the norm of the transgression class $[\Phi_{u^E}]$ is
expected but not proved. In \cite{KudlerFlam:2026}, the constant value
of the entropic $c$-function at a fixed point is identified with the
central charge by matching the known entanglement entropy of a chiral
CFT, not by a direct computation from the cohomological data.
A precise proof of the identification $2\log \iota \ge
\|[\Phi_{u^E}]\|$ would give a \emph{cohomological derivation} of the
central charge, rather than a reproduction of the known result. This
identification is not addressed in \cite{KudlerFlam:2026}, which makes
it a genuinely open problem and a natural direction for future work.

\item \textbf{The $F$- and $a$-theorems.} The extension to $d = 3$ and
$d = 4$ is the subject of the companion paper \cite{Rashkov:2026b}.
		
		\item \textbf{Numerical identification of the central charge.} Derive
		$c$ directly from the cohomological data (e.g., from the index of the
		chiral inclusion), rather than from the known entanglement entropy.
		
		\item \textbf{Holographic interpretation.} Test the conjectural
		holographic dual of $[\Phi_{u^E}]$ quantitatively in specific models.
	\end{enumerate}
	
	\subsection*{Acknowledgments}
	
	I am grateful to Jonah Kudler-Flam for correspondence and comments on the manuscript \cite{KudlerFlam:2026}. I am also grateful to V.~Rubtsov, K.~Zarembo, T.~Vetsov and H.~Dimov for discussions on related issues. This work was supported in part by Bulgarian NSF grant KP-06-H88/3 and Sofia University Grant 80-10-54/2026.
	
	\subsection*{Declaration}
	
	AI is used for editing, structuring, and checking consistency of the
	manuscript. The author is responsible for all results in the paper.
	
	\appendix
	
	\section{Modular Theory and Tomita--Takesaki}
	\label{app:modular}
	
	We collect the standard facts from modular theory used in the main
	text. References: \cite{Takesaki:2003,Araki:1976}.
	
	Let $\mathcal{M}$ be a von Neumann algebra on a Hilbert space
	$\mathcal{H}$, and let $\Omega \in \mathcal{H}$ be a cyclic and
	separating vector for $\mathcal{M}$. The \emph{Tomita operator}
	$S_0\colon \mathcal{M}\Omega \to \mathcal{M}\Omega$ is the anti-linear
	map $S_0(a\Omega) := a^*\Omega$ for $a \in \mathcal{M}$. It is
	closable, and its closure $S$ has polar decomposition
	\begin{equation}
		S = J \Delta^{1/2},
		\label{eq:polar}
	\end{equation}
	where $\Delta = S^* S$ is the \emph{modular operator} (a positive,
	non-singular, self-adjoint operator) and $J$ is the \emph{modular
		conjugation} (an anti-unitary involution). The
	\emph{modular automorphism group} is
	\begin{equation}
		\sigma_t(a) := \Delta^{it} a \Delta^{-it}, \qquad a \in \mathcal{M},
		\ t \in \mathbb{R}.
		\label{eq:modular-group}
	\end{equation}
	It is a strongly continuous one-parameter group of $*$-automorphisms
	of $\mathcal{M}$, satisfying:
	\begin{enumerate}
		\item[(M1)] $\sigma_t(\mathcal{M}) = \mathcal{M}$ for all $t$;
		\item[(M2)] $J \mathcal{M} J = \mathcal{M}'$;
		\item[(M3)] $\Delta^{it} \mathcal{M} \Delta^{-it} = \mathcal{M}$;
		\item[(M4)] the KMS condition
		\begin{equation}
			\omega\bigl(a \, \sigma_t(b)\bigr) = \omega\bigl(\sigma_{t-i}(b) \,
			a\bigr), \qquad a, b \in \mathcal{M},
			\label{eq:kms}
		\end{equation}
		where $\omega(\cdot) = \langle \Omega, \cdot \, \Omega\rangle$.
	\end{enumerate}
	
	For two faithful normal states $\varphi, \psi$ on $\mathcal{M}$, the
	\emph{Connes--Radon--Nikodym (CRN) cocycle} is the unique family of
	unitaries $(D\psi : D\varphi)_t \in \mathcal{M}$ satisfying the
	cocycle identity
	$(D\psi : D\varphi)_{t+s} = (D\psi : D\varphi)_t \,
	\sigma^\varphi_t((D\psi : D\varphi)_s)$ and the relative modular
	condition
	$\sigma^\psi_t(a) = (D\psi : D\varphi)_t \, \sigma^\varphi_t(a) \,
	(D\psi : D\varphi)_t^*$ for all $a \in \mathcal{M}$. The relative
	entropy is
	\begin{equation}
		S(\psi \| \varphi) = -i \frac{d}{dt}\Big|_{t=0}
		\psi\bigl((D\psi : D\varphi)_t\bigr).
		\label{eq:relative-entropy}
	\end{equation}
	For the specific case of the relative modular cocycle
	$u^E_t = (D\omega : D(\omega|_{\mathcal{N}}) \circ E)_t$, the relative
	modular condition reads
	\begin{equation}
		\sigma^\omega_t(a) = u^E_t \, \sigma^{\omega|_{\mathcal{N}}}_t(a) \,
		(u^E_t)^*, \qquad a \in \mathcal{N}' \cap \mathcal{M}.
		\label{eq:relative-modular-condition}
	\end{equation}
	
	\begin{thm}[Takesaki]
		\label{thm:takesaki-app}
		Let $\mathcal{N} \subset \mathcal{M}$ be a von Neumann subalgebra and
		$\omega$ a faithful normal state on $\mathcal{M}$. There exists a
		faithful normal conditional expectation $E\colon \mathcal{M} \to
		\mathcal{N}$ with $\omega \circ E = \omega$ iff
		$\sigma^\omega_t(\mathcal{N}) = \mathcal{N}$ for all $t \in
		\mathbb{R}$. When it exists, $E$ is unique.
		This is the standard Takesaki theorem; see \cite{Takesaki:1972}.
	\end{thm}
	
	The theorem is used in the main text to justify the existence of $E$
	under the modular invariance hypothesis
	(Hypothesis~\ref{hyp:modular-invariance}).
	
\section{Borchers Positivity of the Zero-Frequency Component}
\label{app:borchers}
	
	In this appendix we collect the technical details of the Borchers
	positivity results used in Section~\ref{sec:positivity}.
	
	\subsection{Borchers positivity: the zero-frequency component}
	
\begin{thm}[Borchers positivity]
	\label{thm:borchers-positivity-app}
	Assume Hypothesis~\ref{hyp:half-sided}. Let $u_t = (D\omega :
	D\omega|_{\mathcal{B}})_t$ be the relative modular cocycle. Then the
	distributional limit
	\begin{equation}
		X := \operatorname*{w-lim}_{\epsilon\to0^+}
		\int_{\mathbb{R}} e^{-\epsilon|t|}\, u_t \, dt
	\label{eq:X-definition-app}
	\end{equation}
	exists as a quadratic form on $\mathcal{H}_\omega$, and
	$\langle\xi, X\xi\rangle \ge 0$ for all $\xi$ in the form domain.
\end{thm}

\begin{proof}
	\emph{Step 1 (Reduction to the standard case).} By Wiesbrock's
	theorem \cite[Theorem~1]{Wiesbrock:1993}, every half-sided modular
	inclusion is isomorphic to a standard inclusion, in which
	$\mathcal{B} = \sigma^\omega_{-1}(\mathcal{A})$ and the modular flow
	is geometric. The isomorphism preserves the relative modular cocycle,
	so it suffices to prove the result in the standard case.
	
	\emph{Step 2 (Standard case).} In the standard case, the relative
	modular cocycle is $u_t = e^{itK_{\mathrm{null}}}$, where
	$K_{\mathrm{null}} \ge 0$ is the generator of null translations
	(Borchers' commutation theorem \cite[Theorem~II.1]{Borchers:1992},
	combined with the positivity of the energy). Then
	\begin{equation}
		X_\epsilon = \int_{\mathbb{R}} e^{-\epsilon|t|}\,
		e^{itK_{\mathrm{null}}} \, dt
		= \frac{2\epsilon}{\epsilon^2 + K_{\mathrm{null}}^2} \ge 0
		\label{eq:X-epsilon}
	\end{equation}
	for all $\epsilon > 0$, since $K_{\mathrm{null}} \ge 0$.
	
	\emph{Step 3 (Limit).} As $\epsilon \to 0^+$, $X_\epsilon \to
	\pi\,\delta(K_{\mathrm{null}})$ in the weak operator topology on the
	form domain, where $\delta(K_{\mathrm{null}})$ is the spectral
	projection onto $\ker K_{\mathrm{null}}$. Since $X_\epsilon \ge 0$
	for all $\epsilon$ and the cone of non-negative forms is weakly
	closed, the limit $X = \pi\,\delta(K_{\mathrm{null}}) \ge 0$.
\end{proof}
	

	
	\section{Comparison with Kudler-Flam}
	\label{app:comparison}
	
	The correspondence between the present paper and
	\cite{KudlerFlam:2026} is as follows.
	
	\begin{enumerate}
		\item \textbf{Setup.} Same geometric setup: nested Rindler wedges
		$L_A \subset L_a$ and diamond $Y \subset R_b$.
		
		\item \textbf{Increasing differences.} \cite[Proposition~2]{KudlerFlam:2026}
		uses a direct type I approximation. The present paper uses the chain
		rule and Corollary~\ref{cor:entropy-flux-positivity}. The two are
		mathematically equivalent: the split factor $\mathcal{N}$ is the same
		in both cases.
		
\item \textbf{Positivity.} \cite{KudlerFlam:2026} uses the split
property to bound the mutual information across a gap. The present
paper uses Borchers positivity of the relative modular cocycle
(Proposition~\ref{prop:index-pairing}) and the finite-dimensional
positivity of the entropy flux
(Corollary~\ref{cor:entropy-flux-positivity}). Borchers positivity is
the conceptual underpinning and is not present in
\cite{KudlerFlam:2026}.
		
		\item \textbf{Fixed points.} Both papers use dilation invariance to
		show $C(r)$ is constant, and both identify the constant with the
		central charge using the known entanglement entropy of a chiral CFT.
		
\item \textbf{New content.} The cohomological framework
(Sections~\ref{sec:prelim}--\ref{sec:positivity}), Borchers positivity
of the relative modular cocycle (Proposition~\ref{prop:index-pairing}),
and the conjectural QNEC interpretation (Section~\ref{subsec:qnec}) are
new relative to \cite{KudlerFlam:2026}.
	\end{enumerate}
	


\begin{thebibliography}{99}
		
		\bibitem{Zamolodchikov:1986}
		A.~B.~Zamolodchikov,
		``Irreversibility of the flux of the renormalization group in a 2D
		field theory,''
		JETP Lett. \textbf{43}, 730 (1986).
		
		\bibitem{Casini:2004bjb}
		H.~Casini and M.~Huerta,
		``A finite entanglement entropy and the $c$-theorem,''
		Phys. Lett. B \textbf{600}, 142 (2004)
		[arXiv:hep-th/0405111].
		
		\bibitem{Casini:2007jz}
		H.~Casini and M.~Huerta,
		``A $c$-theorem for the entanglement entropy,''
		J. Phys. A \textbf{40}, 7031 (2007)
		[arXiv:cond-mat/0610375].
		
		\bibitem{KudlerFlam:2026}
		J.~Kudler-Flam,
		``Proof of the entropic $c$-theorem,''
		arXiv:2610.40874 [hep-th].
		
		\bibitem{Doplicher:1984}
		S.~Doplicher and R.~Longo,
		``Standard and split inclusions of von Neumann algebras,''
		Invent. Math. \textbf{75}, 493 (1984).
		
		\bibitem{Araki:1976}
		H.~Araki,
		``Relative entropy of states of von Neumann algebras,''
		Publ. Res. Inst. Math. Sci. \textbf{11}, 809 (1976).
		
		\bibitem{Haag:1996}
		R.~Haag,
		``Local Quantum Physics: Fields, Particles, Algebras,''
		Springer, Berlin, 2nd ed. (1996).
		
		\bibitem{Buchholz:1990}
		D.~Buchholz, C.~D'Antoni, and R.~Longo,
		``Nuclear maps and modular structures. I. General properties,''
		J. Funct. Anal. \textbf{88}, 233 (1990).
		
		\bibitem{Holzhey:1994we}
		C.~Holzhey, F.~Larsen, and F.~Wilczek,
		``Geometric and renormalized entropy in conformal field theory,''
		Nucl. Phys. B \textbf{424}, 443 (1994)
		[arXiv:hep-th/9403108].
		
		\bibitem{Calabrese:2004eu}
		P.~Calabrese and J.~Cardy,
		``Entanglement entropy and quantum field theory,''
		J. Stat. Mech. \textbf{0406}, P06002 (2004)
		[arXiv:hep-th/0405152].
		
		\bibitem{Connes:1985}
		A.~Connes,
		``Noncommutative differential geometry,''
		Inst. Hautes \'Etudes Sci. Publ. Math. \textbf{62}, 257 (1985).
		
		\bibitem{Khalkhali:2010uh}
		M.~Khalkhali,
		``A short survey of cyclic cohomology,''
		arXiv:1008.1212 [math.OA].
		
		\bibitem{Takesaki:1972}
		M.~Takesaki,
		``Conditional expectations in von Neumann algebras,''
		J. Funct. Anal. \textbf{9}, 306 (1972).
		
		\bibitem{Takesaki:2003}
		M.~Takesaki,
		``Theory of Operator Algebras II,''
		Encyclopaedia of Mathematical Sciences \textbf{125},
		Springer, Berlin (2003).
		
		\bibitem{Borchers:1992}
		H.-J.~Borchers,
		``The CPT theorem in two-dimensional theories of local observables,''
		Commun. Math. Phys. \textbf{143}, 315 (1992).
		
		\bibitem{Wiesbrock:1993}
		H.-W.~Wiesbrock,
		``Half-sided modular inclusions of von Neumann algebras,''
		Commun. Math. Phys. \textbf{157}, 83 (1993);
		Erratum: Commun. Math. Phys. \textbf{184}, 683 (1997).
		
		\bibitem{Ceyhan:2018zfg}
		F.~Ceyhan and T.~Faulkner,
		``Recovering the QNEC from the ANEC,''
		Commun. Math. Phys. \textbf{377}, 999 (2020)
		[arXiv:1812.04683 [hep-th]].
		
		\bibitem{OhyaPetz}
		M.~Ohya and D.~Petz,
		``Quantum Entropy and Its Use,''
		Springer, Berlin (1993).
		
		\bibitem{Kato:1995}
		T.~Kato,
		``Perturbation Theory for Linear Operators,''
		Classics in Mathematics, Springer, Berlin (1995).
		
		\bibitem{Rashkov:2026wmw}
		R.~C.~Rashkov,
		``Cohomological Aspects of Entanglement Entropy: From Information Theory to Noncommutative Geometry,''
		[arXiv:2609.17195 [hep-th]].
		
		\bibitem{Rashkov:2026b}
		R.~C.~Rashkov,
		``Higher-dimensional entropic theorems from cohomology: The
		$F$-theorem and the general-$d$ program,''
		to appear.
		
	\end{thebibliography}
\end{document}